\documentclass[aps,prl,twocolumn,showpacs,amsmath,amssymb,footinbib]{revtex4}
\usepackage{graphicx}
\usepackage{amsmath}
\usepackage{enumitem}
\usepackage{natbib}
\usepackage{float}
\usepackage{verbatim}
\usepackage[abs]{overpic}

\begin{document}

\title{Control of Andreev bound state population and related charge-imbalance effect}
\author{Roman-Pascal~Riwar, Manuel Houzet, Julia S. Meyer}
\affiliation{Univ. Grenoble Alpes, INAC-SPSMS, F-38000 Grenoble, France.\\
CEA, INAC-SPSMS, F-38000 Grenoble, France.
}
\author{Yuli~V.~Nazarov}
\affiliation{Kavli Institute of NanoScience, Delft University of Technology, Lorentzweg 1, NL-2628 CJ, Delft, The Netherlands.}

\begin{abstract}
Motivated by recent experimental research, we study the processes in an ac~driven superconducting constriction whereby one quasiparticle is promoted to the delocalized states outside the superconducting gap. We demonstrate that with these processes one can control the population of the Andreev bound states in the constriction. 
We stress an interesting charge asymmetry of these processes that may produce a charge imbalance of accumulated quasiparticles, which depends on the phase.

\end{abstract}


\maketitle

Superconducting nanodevices are among the most promising candidates to realize quantum computation in the solid state~\cite{Devoret2013}, and for many other applications. Quasiparticle poisoning, whereby an unwanted quasiparticle enters a bound state in the device, is an important factor harming their proper operation~\cite{Catelani2011}. Naively, the superconducting gap $\Delta$ should ensure an exponentially small quasiparticle concentration at low temperatures. However, various experiments indicate that a long-lived, non-equilibrium quasiparticle population persists in the superconductor, affecting the operation of various superconducting devices~\cite{Martinis2009,Lenander2011,Rajauria2012,Riste2013,Wenner2013,LevensonFalk2014}, 
including tempting proposals to use Majorana states in superconductors~\cite{Fu2009,vanHeck2011,Rainis2012}.

This makes it important to develop the means of an active control of the quasiparticle population in bound states associated with a nano-device.

As a generic model we consider a superconducting constriction with a few highly transparent channels. Such constrictions are made on the basis of atomic break junctions~\cite{SupQPC}. The simplicity of their theoretical description enabled detailed theoretical research~\cite{Virtanen2010,Kos2013,Olivares2014}.  In the presence of a phase difference at the constriction, an Andreev bound state (ABS) is formed in each channel~\cite{Andreev1964,QuantumTransport}. In a recent pioneering experiment~\cite{BretheauRPL}, the population of such a single bound state has been detected by its effect on the supercurrent in the constriction. The spectroscopy of Andreev states has also been  successfully performed in this setup~\cite{Bretheau2013,Bretheau2013b}. Thus motivated, we theoretically investigate the control of the population of quasiparticles in the ABS at a superconducting constriction by means of pulses of high-frequency microwave irradiation.

\begin{figure}
\includegraphics[scale=0.9]{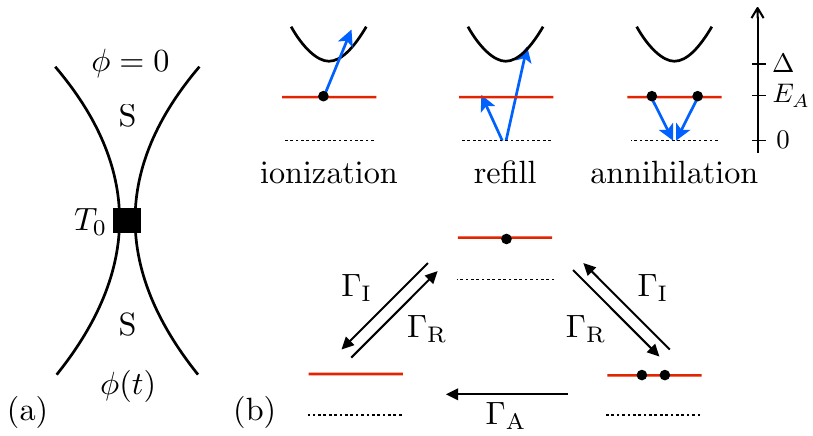}
\caption{Shown in (a) is the setup. A single-channel superconducting constriction with transmission coefficient $T_0$, biased by the phase difference $\phi(t)$. In (b) we depict the processes changing the population of the ABS and the transitions they induce.}\label{fig_setup_processes}
\end{figure}

In this Letter, we demonstrate that an efficient control of the ABS can be achieved by inducing the processes of \textit{ionization} and \textit{refill} (Fig. \ref{fig_setup_processes}), due to an ac modulation of the phase drop across the junction, $\phi(t)=\phi+\delta\phi\sin(\Omega t)$. In the course of such a process, a quasiparticle is promoted to the delocalized states and leaves the constriction. We compute the rates of these processes in the lowest order in irradiation amplitude $\delta \phi$. We find an interesting charge asymmetry of the emitted quasiparticles. This asymmetry leads to a net quasiparticle current and charge imbalance of the quasiparticles accumulated in the vicinity of the constriction. Charge imbalance can be measured by a standard setup using a normal-superconducting (N-S) tunnel junction ~\cite{Langenberg1986,Tinkham1972b,Tinkham1972,Hubler2010,Golikova2014}.
 
We focus on the regime of low temperatures which permits to neglect the equilibrium population of delocalized quasiparticle states. Let us consider a quasiparticle in the ABS with energy $E_A <\Delta$. If we modulate the superconducting phase with the frequency $\Omega> \Delta - E_A$  ($\hbar=1$), we can transfer this quasiparticle to the states of the delocalized spectrum. This is an {\it ionization} process. Suppose we start with no quasiparticle in the constriction and wish to fill the bound state. This can be achieved by the absorption of a quantum of the high-frequency phase modulation, provided the energy quantum exceeds $\Delta+E_A$. In the course of such a {\it refill} process, one quasiparticle  emerges in the Andreev level while another one is promoted to the delocalized states and leaves the constriction.

We model the constriction with an effective 1D Hamiltonian~\cite{supplink}. The advantage of the model in use is that we can express all the characteristics of the bound state and transition dynamics with a single transmission coefficient $T_0$ characterizing the channel.  For instance, the energy of the spin-degenerate ABS reads $E_A= \Delta\sqrt{1-T_0\sin^2(\phi/2)}$.

\begin{figure}[tb]
\includegraphics[scale=1.1]{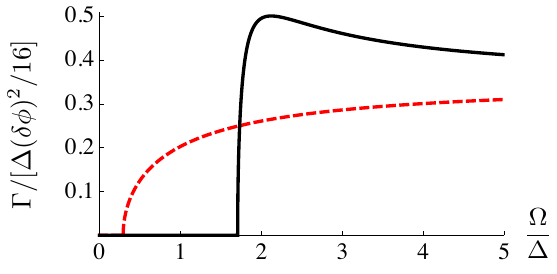}
\caption{Ionization (dashed) and refill (solid) rates for $T_0=0.5$ and $\phi=\pi$, when $E_A \approx 0.7 \Delta$. The ionization rate appears at the threshold $\Omega \approx 0.3 \Delta$, while the threshold for the refill is $\approx 1.7 \Delta$.  }\label{fig_rates}
\end{figure}

The explicit expressions for the rates in lowest order in the phase modulation amplitude read as follows,
\begin{widetext}
\begin{equation}\label{eq_Gamma_I_R}
\Gamma_\text{I,R} = \frac{T_0 (\delta\phi)^2}{16} \theta(\Omega \pm E_A - \Delta) \frac{\sqrt{\Delta^2-E^2_A}}{E_A}\sqrt{(\Omega \pm E_A)^2-\Delta^2} \frac{E_A\Omega \pm \Delta^2(\cos\phi+1)}{(\Omega\pm E_A)^2-E^2_A}\ .
\end{equation}
\end{widetext}
We see that the ionization and refill rates at $T_0 \simeq 1$ are of the order of $(\delta \phi)^2 \Delta$ and, at sufficiently large phase modulation amplitudes, are restricted by $\Delta$ only. Thus the population of the ABS can be changed quickly. 
We illustrate the frequency dependence of the ionization and refill rates in Fig. \ref{fig_rates}. In the limit of large frequencies, both rates saturate at the same value. We stress, however, that the practical frequencies for the manipulation of the ABS are most likely restricted by $2 \Delta$: higher frequencies would cause massive generation of quasiparticle pairs at the constriction and in the bulk of the superconductor.

In addition to the microwave-induced processes, there are intrinsic processes changing the ABS occupation. For low temperatures, the only such process is the annihilation of two quasiparticles in the same ABS. This inelastic process
is due to quantum fluctuations of the phase and is expressed in terms of the phase noise spectrum $S_{\phi}(\omega)$ that is related to the impedance $Z(\omega)$ of the external circuit felt by the constriction. Namely, 
$S_\phi(\omega)=4\pi G_Q Z(\omega)/\omega$, where $\omega>0$, $T\ll \omega$, and $G_Q \equiv e^2/\pi\hbar$, while
\begin{equation}\label{eq_Gamma_A}
\Gamma_\text{A} = \frac{S_{\phi}(2E_A)}{4} \left( 1- \frac{E^2_A}{\Delta^2}\right) \left(\Delta^2 -E^2_A- 4 \left(\frac{\partial E_A}{\partial \phi}\right)^2\right)\ .
\end{equation}
It may be estimated as $\Gamma_\text{A} \simeq \langle\langle \phi^2\rangle\rangle_q \Delta$, $\langle\langle \phi^2\rangle\rangle_q \simeq G_Q Z $ being the quantum fluctuation of the phase. For typical electromagnetic environments, $Z$ is of the order of the vacuum impedance and $\langle\langle \phi^2\rangle\rangle_q \simeq 10^{-3}$. Thus, by controlling the ac amplitude with respect to the phase noise, both regimes of negligible, $(\delta\phi)^2 \gg \langle\langle \phi^2\rangle\rangle_q$, and fast, $(\delta\phi)^2 \ll \langle\langle \phi^2\rangle\rangle_q$, annihilation are in principle reachable. We discuss both limiting cases in the following.

With the rates \eqref{eq_Gamma_I_R}~and~\eqref{eq_Gamma_A} we can determine the distribution of the bound state populations under constant driving. The processes causing transitions between the ABS occupations $n=0,1,2$ are summarized in Fig.~\ref{fig_setup_processes}.  
The master equation for the probabilities $P_n$ reads
\begin{subequations}\label{eq_master}
\begin{align}
\dot{P}_0&=-2\Gamma_\text{R} P_0+\Gamma_\text{I} P_1+\Gamma_\text{A} P_2,\\
\dot{P}_1&=-(\Gamma_\text{I}+\Gamma_\text{R})P_1 +2\Gamma_\text{R} P_0+2\Gamma_\text{I} P_2,\\
\dot{P}_2&= - (\Gamma_\text{A}+2\Gamma_\text{I})P_2 +\Gamma_\text{R} P_1 .
\end{align}
\end{subequations}
The factors $2$ in this equation are due to the spin degeneracy of the single quasiparticle state.

In the absence of a refill rate, $\Gamma_\text{R} =0, \Gamma_\text{I} \ne 0$, the ABS is always emptied by the ionization processes, $P_0^\text{st}=1$. Therefore the ac phase modulation can be used for \lq purification\rq ~of the localized quasiparticle states in  nanodevices.
We stress that the opposite situation, $\Gamma_\text{I} =0, \Gamma_\text{R} \ne 0$, is not achievable since the phase modulation responsible for refill processes also produces ionization. In this case, the constant ac modulation will cause a random distribution of the population~\cite{supplink}.

An efficient manipulation of the population is yet possible, provided one can measure the result of the manipulation, that is, the population. This is equivalent to measuring the superconducting current response of the constriction. A practical measurement would most likely address the inductive response that takes three discrete values following the population of the state, $I_n=I_A(1-n)$, where $I_A=-2e\partial_\phi E_A$. 

First, let us concentrate on a simple situation when the annihilation rate is significant at the time span of the measurement and manipulation. In this case, the doubly occupied state is unstable and only $n=0,1$ are achievable. This corresponds to the current experimental situation ~\cite{BretheauRPL}. If $n=1$ and we wish to set $n=0$, we just need to apply an ac~pulse with the frequency $E_A-\Delta <\Omega<E_A+\Delta$ and a duration exceeding $\Gamma_\text{I}^{-1}$. If $n=0$ and the desired state is $n=1$, we will apply a refill pulse and {\it measure} the result. If $n=1$, we are there. If not, we apply another pulse. 

Since the ac manipulation is fast, it is plausible to control the population even at time scales $\simeq \Gamma_\text{A}$ with a similar scheme and go from any $n=0,1,2$ to any $m=0,1,2$, combining measurement as well as refill and ionization pulses. Naturally, this requires the measurement time to be much shorter than $\Gamma_\text{A}^{-1}$. If $n>m$, we apply ionization pulses, otherwise refill pulses permitting occasional ionization.
The frequency and duration of the pulse can be optimized to boost the rate and the probability to come to the desired state with a minimum number of measurements. For instance, for $n=2,m=1$, the optimal duration of the ionization pulse is $\Gamma_\text{I}^{-1}\ln2$ that results in the maximum 50\% probability to achieve $m=1$ with a single pulse.

We find a very interesting asymmetry of the quasiparticles emitted in the course of the described processes. While the quasiparticles fly with equal probability to both leads, more electron-like quasiparticles leave to one of the leads while more hole-like ones leave to the opposite lead. This results in a net {\it charge transfer} per process that we define as $q_\alpha(E)=\frac{\sqrt{E^2-\Delta^2}}{E}\frac{\Gamma_{\alpha\text{e}}-\Gamma_{\alpha\text{h}}}{\Gamma_{\alpha\text{e}}+\Gamma_{\alpha\text{h}}}$, where $E$ is the energy of the emitted quasiparticle and $\alpha=\text{I,R}$. The rates $\Gamma_{\alpha\text{e}}$ and $\Gamma_{\alpha\text{h}}$ are the partial rates for electron- and hole-like quasiparticles, respectively~\cite{supplink}. In the following we choose to focus on the charge transfer to the right electrode.

\begin{figure}
\centering
\includegraphics[scale=1]{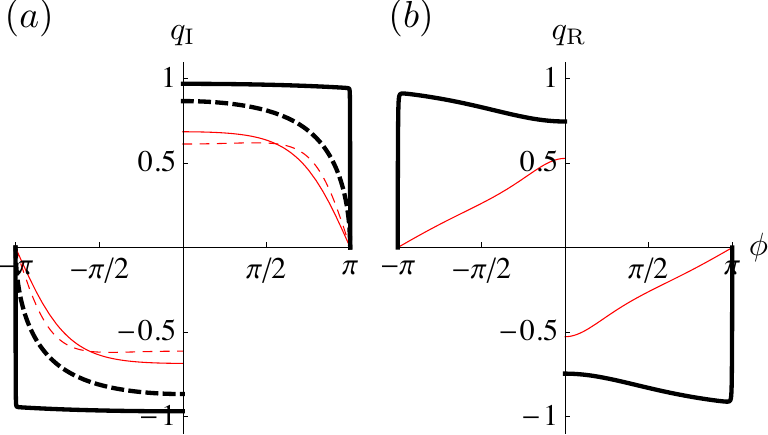}
\caption{The charge transfers due to (a) ionization, $q_\text{I}$, and (b) refill, $q_\text{R}$, as a function of $\phi$. The parameters are $\Omega/\Delta=1$ (dashed) and $T_0=\{0.5,1\}$ (thin and thick) as well as $\Omega/\Delta=3$ (solid) $T_0=\{0.5,1\}$. Note that $q_\text{R}=0$ for $\Omega/\Delta=1$.}\label{fig_pR}
\end{figure}

Upon evaluating the rates, the charge transfers $q_\text{I,R}$ for the processes considered are expressed as
\begin{align}\label{eq_qIR}
q_\text{I,R}&=\mp2\frac{\partial E_A}{\partial \phi}\sqrt{\frac{(\Omega\pm E_A)^2-\Delta^2}{\Delta^2-E_A^2}}\frac{E_A\left(1+\frac{E_A}{\Omega\pm E_A}\right)}{\Omega E_A\pm\Delta^2\left(1+\cos\phi\right)}\ .
\end{align}
In Fig.~\ref{fig_pR}, $q_\text{I,R}$ as a function of $\phi$ are plotted for several parameters. We see immediately that $q_\alpha(\phi)=-q_\alpha(-\phi)$, like the supercurrent. 
Inverting the phase therefore inverts the charge transfer.

Contrary to the supercurrent,  the charge transfer exhibits a discontinuity at $\phi=0$. The explanation of this rather counterintuitive feature is that the wave function of the Andreev bound state is not a continuous function of $\phi$ at $\phi=0$, since the state merges with the delocalized spectrum at this point.  The charge transfers $q$ are $2\pi$-periodic and have a node at $\phi=\pi$, where the charge asymmetry vanishes. In addition, $q_\text{I}$ and $q_\text{R}$ are generally of opposite sign. The maximum charge transfer for a given $\phi$ is reached in the limit of a fully transparent constriction $T_0 \to 1$ (thick curves in Fig.~\ref{fig_pR}) where the ac drive actually produces only a quasiparticle of one kind, namely, e-like (h-like) for $0<\phi<\pi$ ($-\pi<\phi<0$). 

Under constant irradiation the net charge transfer per unit time is computed from the master equation~\eqref{eq_master} and reads
\begin{equation}
\label{eq:dotq}
\dot{q} = q_\text{I}\Gamma_\text{I}(P_1 +2 P_2) + q_\text{R}\Gamma_\text{R}(2 P_0 + P_1)\ .
\end{equation}
We see that the refill process is crucial for the net effect: otherwise the ABS will always be empty (expressions for $\dot{q}$ in in the limits of fast and slow annihilation are provided in~\cite{supplink}).

\begin{figure}
\includegraphics[scale=0.9]{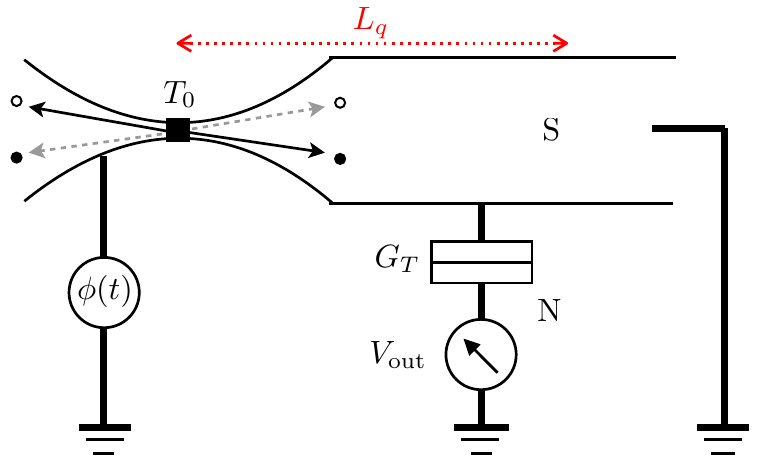}
\caption{ Build-up of charge imbalance due to charge asymmetry of the quasiparticles emitted from the constriction. The solid and dashed arrows indicate the dominant processes and their weaker counterparts, respectively. The filled (empty) circles represent electron-like (hole-like) quasiparticles. The charge imbalance is measured with a N-S tunnel junction probe attached to the lead, with a tunnel conductance $G_T$. The voltage $V_{\text{out}}$ is the output signal for the voltage measurement. }\label{fig_charge_imbalance_sketch}
\end{figure}

If the thermalization of the  quasiparticle distribution in the leads near the constriction is not immediate, the effect leads to charge imbalance ~\cite{Langenberg1986,Tinkham1972b}. Namely, the charge transfer asymmetry gives rise to the build-up of a non-equilibrium quasiparticle charge density $\rho$. This charge imbalance can be measured with a normal-metal voltage probe attached to the superconductor: the method proposed in \cite{Tinkham1972} and widely applied in recent years ~\cite{Hubler2010,Golikova2014}, see Fig. \ref{fig_charge_imbalance_sketch}. In this case, $\rho$ gives rise to a current $I_q$ at the N-S tunnel junction. Applying a voltage $eV_{\text{out}}=\mu_\text{N}-\mu_\text{S}$ between the normal metal and superconducting contacts produces a counter-current $I_V$. The voltage $V_{\text{out}}$ at which $I_q+I_V=0$ is the signal of the charge imbalance.

For $T\ll\Delta$, this measurement is extremely sensitive owing to the fact that
$I_V$ is formed by the normal-metal excitations with energies $>\Delta$. Since at low temperatures the number of these excitations is exponentially small, an exponentially large $V_{\text{out}}$ is required to compensate $I_q$. In the linear regime, the signal voltage 
reads $eV_{\text{out}} = T \rho/c_0$, where
$c_0 = \nu_0 \sqrt{2\pi T\Delta}e^{-\Delta/T}$ is the equilibrium quasiparticle density and $\nu_0$ the normal metal density of states. Owing to this, even at moderately low $T = 0.05 \Delta$ in aluminum, a charge imbalance of 0.001 elementary charges per cubic micrometer produces already a signal $\simeq 0.1 T/e$.
The above relation is valid if $eV_{\text{out}} \ll T$, at larger imbalances the signal saturates at
$T \ln(\rho/c_0)$~\cite{supplink}.

To estimate $\rho$, we note that  potential scattering does not lead to the relaxation of charge imbalance: this requires  inelastic processes and/or scattering on magnetic impurities~\cite{Tinkham1972}.   The charge imbalance lifetime $\tau_q$ is therefore long and quasiparticles diffuse far away from the constriction, spreading over the length scale $L_q \simeq \sqrt{D\tau_q}$, $D$ being the quasiparticle diffusion coefficient. We assume the N-S voltage probe to be placed within this scale.
The created quasiparticles are distributed over $\mathcal{V}$, the volume of the lead at the scale $L_q$.
We note that the normal-state resistance of this piece of the lead can be estimated as $R_q^{-1} = e^2 \nu_0 D \mathcal{V}/L_q^2$. This permits to represent the estimation in a compact form~\cite{supplink}, independent of peculiarities of the geometry and disorder in the leads: 
$\rho \simeq (R_q G_Q) \nu_0 \dot{q}$.

Combining estimations for $V_{\text{out}}$ and $\rho$, and estimating ${\dot q} \simeq(\delta\phi)^2\Delta\equiv\Gamma$, we find
\begin{equation}
eV_{\text{out}}  \simeq (R_q G_Q) T\frac{\nu_0}{c_0} \Gamma \simeq (R_q G_Q) \sqrt{\frac{T}{\Delta}}e^{\Delta/T}\Gamma\ .
\end{equation}
To get a rough estimate of achievable values, we  take $R_q \simeq 1\,\text{Ohm}$, $\Gamma \simeq 10^{-3} \Delta \simeq 1 ~\mu\text{eV}$, $T \simeq 0.05 \Delta$. Without the exponential factor, the value of $V_{\text{out}}$ would be in the nanovolt range. However, the exponential factor yields nine orders of magnitude. Since such an estimation greatly exceeds $T$, the signal voltage in this case would saturate at the value $T \simeq 10 ~\mu\text{eV}$ which is easy to measure.

An alternative measurement is to use a grounded N-S junction. The current signal would then be due to the emitted quasiparticles slipping to the normal electrode. If the junction conductance $G_T$ is sufficiently large, $G_T R_q\gg1$, all emitted quasiparticles would do so resulting in $I_{\text{out}} = e \dot{q} \simeq e \Gamma$. 

We studied the processes of quasiparticle emission in a superconducting constriction subject to an ac phase modulation and proposed an efficient scheme to control the occupation of the ABS. In addition, we found an asymmetry of the rates of electron- and hole-like quasiparticle emission. This asymmetry is expected to lead to a charge imbalance of the quasiparticles accumulated near the constriction which may be measured in an open or closed circuit geometry. Our results may be generalized to the multi-channel case, by summing up the contributions of each channel.

This work has been supported by the Nanosciences Foundation in Grenoble, in the frame of its Chair of Excellence program. MH and JSM acknowledge support through grants No. ANR-11-JS04-003-01 and No. ANR-12-BS04-0016-03, and an EU-FP7 Marie Curie IRG. One of the authors (RPR) would like to thank Tatiana Krishtop and Lars Elster for stimulating discussions.

\bibliographystyle{apsrev}
\bibliography{bib_superconducting}

\begin{thebibliography}{3}
\bibitem[{foo()}]{footsupp}
\bibinfo{note}{The expression for $\Gamma_\text{R}$ may be readily understood
  in the semiconductor picture where the refill process corresponds to
  transferring a quasiparticle from a delocalized state at negative energies to
  the ABS.}
\end{thebibliography}

\end{document}


\setcounter{page}{6}

\title{SUPPLEMENTARY MATERIAL:\\ Control of Andreev bound state population and related charge-imbalance effect}
\author{Roman-Pascal~Riwar, Manuel Houzet, Julia S. Meyer}
\affiliation{Univ. Grenoble Alpes, INAC-SPSMS, F-38000 Grenoble, France.\\
CEA, INAC-SPSMS, F-38000 Grenoble, France.
}
\author{Yuli~V.~Nazarov}
\affiliation{Kavli Institute of NanoScience, Delft University of Technology, Lorentzweg 1, NL-2628 CJ, Delft, The Netherlands.}

\maketitle

\section{1D model of a single-channel superconducting constriction}
\label{app:model}
We model the superconducting weak link with a 1D quantum Hamiltonian corresponding to a single transport channel. The constriction of length $L$ is modeled by a scattering potential $V(x)$, with the spatial coordinate $x$. A finite  vector potential $A(x)$ on a local support provides a phase bias between the left and right contact, $\phi=2e\int dx A(x)$.
We focus on the regime where the excitation energy is much smaller than the Fermi energy, such that the spectrum can be linearized. Left/right moving electrons with the Fermi wave vector $\mp k_F$ are represented by a pseudo spin vector basis, $|\mp\rangle$, where $\sigma_z=|-\rangle\langle-|-|+\rangle\langle+|$. In the linearized regime, the current density operator is represented as $j=-ev_F\sigma_z$. The Bogoliubov-de Gennes Hamiltonian is then given as
\begin{equation}\label{eq_Hamiltonian}
H=\left[-iv_F\partial_x\sigma_z+V(x)\sigma_x\right]\tau_z\\-ev_FA(x)\sigma_z+\Delta\tau_x\ ,
\end{equation}
where the Pauli matrices $\tau_i$ represent the Nambu space and $v_F$ is the Fermi velocity.
The potential $V(x)\sigma_x$ provides a finite reflection probability from left to right movers and vice versa, with $\sigma_x=|-\rangle\langle+|+|+\rangle\langle-|$. The potential function $V(x)$ is a real function and has a finite support in the interval $x\in[0,L]$.

\section{Diagonalization at stationary phase}

First, we diagonalize the Hamiltonian~\eqref{eq_Hamiltonian} for a stationary phase $\phi$. Assuming a short constriction, $L\ll v_F/\Delta$, there is one Andreev bound state solution $|\varphi_A(x)\rangle$ with a subgap eigenenergy $E_A=\Delta\sqrt{1-T_0\sin^2(\phi/2)}$. The normal state transmission coefficient $T_0$  characterizes the transport channel under consideration. The Andreev bound state is responsible for the supercurrent in the constriction. In addition, there are the extended scattering eigenstates $|\varphi_{\beta\eta}^{\text{out}}(x)\rangle$ with eigenenergies $E>\Delta$, where an $\eta$-like quasiparticle ($\eta=\text{e,h}$) is outgoing to the left/right contact, $\beta=\text{l,r}$. They have the BCS density of states $\nu(E)=\theta(E-\Delta)E/\sqrt{E^2-\Delta^2}\nu_0$, where $\nu_0$ is the normal metal density of states. This set of outgoing states is connected to the incoming scattering states via the scattering matrix $S_{\beta\eta}^{\beta'\eta'}=\langle\varphi^{\text{out}}_{\beta'\eta'}|\varphi^{\text{in}}_{\beta\eta}\rangle$. Our scattering matrix coincides with the one found in Ref.~\cite{Olivares2014}. 

\section{Perturbation theory}
\label{app:PT}

We treat the ac drive of the phase, $\delta\phi\sin(\Omega t)$, as a perturbation, and we compute the rates of various processes in the lowest order, $\sim(\delta\phi)^2$, applying Fermi's golden rule. The advantage of the model and the gauge in use is that the matrix elements of the perturbation only depend on the wave functions $|\varphi(x)\rangle$ at the origin.
The rate of ionization from the bound state outgoing to a delocalized $\eta$-like quasiparticle state outgoing to contact $\beta$ with energy $E=\Omega+E_A$ reads 
\begin{equation}
\label{GammaI}
\Gamma_{A\rightarrow \beta\eta}=\frac{\pi}{8}(\delta\phi)^2\nu(E)\left|\langle\varphi_{A}(0)|j|\varphi_{\beta\eta}^\text{out}(0)\rangle\right|^2\ .
\end{equation}
The rate of the refill process whereby a Cooper pair is broken and the quasiparticles occur in the bound state and in the continuum, at energy $E=\Omega-E_A$, reads
\begin{equation}
\label{GammaR}
\Gamma_{0\rightarrow A \beta\eta}=\frac{\pi}{8}(\delta\phi)^2\nu(E)\left|\langle\varphi_{\beta\eta}^\text{out}(0)|j|\tilde{\varphi}_A(0)\rangle\right|^2\ ,
\end{equation}
with $|\tilde{\varphi}_A(0)\rangle=i\tau_y\sigma_x|\varphi_A(0)\rangle^*$~\cite{footsupp}. The partial rates used in the main text are defined as $\Gamma_{\text{I}\eta}\equiv\Gamma_{A\rightarrow\text{r}\eta}$ and $\Gamma_{\text{R}\eta}\equiv\Gamma_{0\rightarrow A\text{r}\eta}$, as we compute the charge imbalance on the right hand side of the constriction. Note that due to charge conservation the corresponding charge transfer on the left side is simply opposite. The total ionization and refill rates are defined as $\Gamma_\text{I}\equiv\sum_{\beta,\eta}\Gamma_{A\rightarrow\beta\eta}$ and $\Gamma_\text{R}\equiv\sum_{\beta,\eta}\Gamma_{0\rightarrow A\beta\eta}$.

In addition, we include quantum phase fluctuations, such that the phase modulation becomes an operator, $\delta\phi(t)\rightarrow\hat{\phi}$, whose dynamics is determined by the electromagnetic environment of the junction.
The phase noise spectrum is $S_{\phi}(\omega)=\int dt\ e^{-i\omega t}\langle\delta\hat{\phi}\left(0\right)\delta\hat{\phi}^{\dagger}\left(t\right)\rangle_\text{env}$, where the expectation value is taken with respect to the environment degrees of freedom. If the environment is in thermal equilibrium, the noise can be related to the impedance $Z(\omega)$ felt by the constriction via the fluctuation dissipation theorem,
$S(\omega)=4\pi G_Q Z(\omega)/\omega$.
The corresponding rate is computed as
\begin{equation}
\label{GammaA}
\Gamma_\text{A} \equiv \Gamma_{2A\rightarrow0}^\text{fluct.}=S_\phi\left(2E_A\right)\left|\langle\varphi_A|j|\tilde{\varphi}_A\rangle\right|^2\ ,
\end{equation}
which results in Eq.~(2) in the main text.

The stationary occupation probabilities of the ABS due to these rates are given by
\begin{subequations}
\begin{align}
{P}_0^\text{st}&=\frac{2\Gamma_\text{I}^2+\Gamma_\text{A}(\Gamma_\text{I}+\Gamma_\text{R})}{\Gamma_\text{I}(\Gamma_\text{A}+2\Gamma_\text{I}+4\Gamma_\text{R})+\Gamma_\text{R}(2\Gamma_\text{R}+3\Gamma_\text{A})}\ ,\\
{P}_1^\text{st}&=\frac{2\Gamma_\text{R}(\Gamma_\text{A}+2\Gamma_\text{I})}{\Gamma_\text{I}(\Gamma_\text{A}+2\Gamma_\text{I}+4\Gamma_\text{R})+\Gamma_\text{R}(2\Gamma_\text{R}+3\Gamma_\text{A})}\ ,
\end{align}
\end{subequations}
with $P_0+P_1+P_2=1$.

\begin{figure}
\includegraphics[scale=0.8]{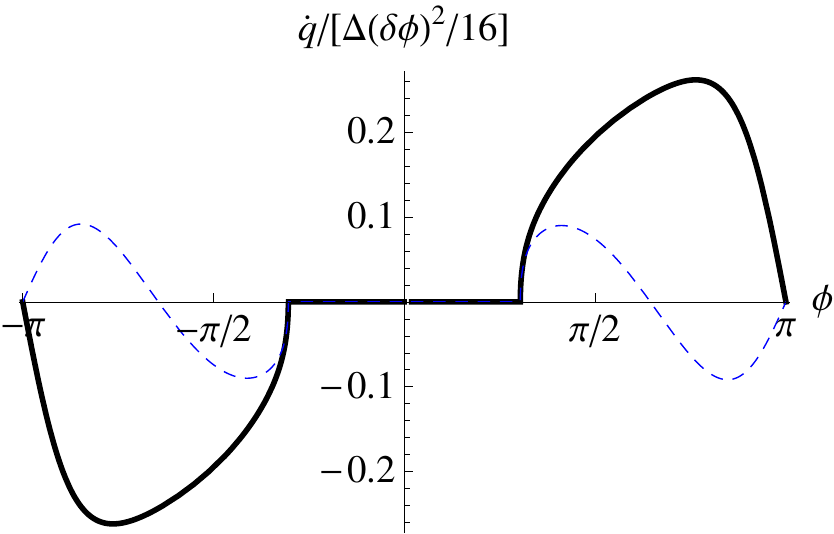}
\caption{The net charge $\dot{q}$ as a function of $\phi$ for the two limiting cases $\Gamma_\text{A}\ll\Gamma_\text{R,I}$ (solid) and $\Gamma_\text{A}\gg\Gamma_\text{R,I}$ (dashed). The parameters are $T_0=0.9$ and $\Omega=1.9\Delta$.}\label{fig_qdot}
\end{figure}

In the following we provide the limits of slow and fast annihilation, $\Gamma_\text{A}\ll\Gamma_\text{I,R}$ and $\Gamma_\text{A}\gg\Gamma_\text{I,R}$, respectively, for a discussion of $\dot{q}$. When $\Gamma_\text{A}$ is small, we find that $P_0^\text{st}=\Gamma_\text{I}^2/(\Gamma_\text{I}+\Gamma_\text{R})^2$, $P_1^\text{st}=2\Gamma_\text{I}\Gamma_\text{R}/(\Gamma_\text{I}+\Gamma_\text{R})^2$, and $P_2^\text{st}=\Gamma_\text{R}^2/(\Gamma_\text{I}+\Gamma_\text{R})^2$. The resulting net charge transfer is then given as
\begin{equation}\label{eq_qdot_no_annihil}
\dot{q}=\left(q_\text{I}+q_\text{R}\right)\frac{2\Gamma_\text{I}\Gamma_\text{R}}{\Gamma_\text{I}+\Gamma_\text{R}}\ .
\end{equation}
Note that that $|q_\text{I}|\geq|q_\text{R}|$ and thus, for slow annihilation the charge transfer due to the ionization process is always dominant (i.e., $\dot{q}$ has the same sign as $q_\text{I}$), see also Fig.~\ref{fig_qdot}.

The stationary probabilities in the opposite limit of fast $\Gamma_\text{A}$ are $P_0^\text{st}=(\Gamma_\text{I}+\Gamma_\text{R})/(\Gamma_\text{I}+3\Gamma_\text{R})$, $P_1^\text{st}=2\Gamma_\text{R}/(\Gamma_\text{I}+3\Gamma_\text{R})$, and $P_2^\text{st}=0$. Here we find
\begin{equation}\label{eq_qdot_fast_annihil}
\dot{q}=\left(q_\text{I}+q_\text{R}\right)\frac{2\Gamma_\text{I}\Gamma_\text{R}}{\Gamma_\text{I}+3\Gamma_\text{R}}+q_\text{R}\frac{2\Gamma_\text{R}^2}{\Gamma_\text{I}+3\Gamma_\text{R}}\ .
\end{equation}
While the first term has the same sign as the expression in Eq.~\eqref{eq_qdot_no_annihil}, the second term can give rise to the change of sign in the charge transfer as shown in Fig.~\ref{fig_qdot}. This is because a fast $\Gamma_\text{A}$ suppresses the two quasiparticle state, $P_2\rightarrow 0$, such that the refill process can become dominant, signified by the extra term in Eq.~\eqref{eq_qdot_fast_annihil}.
In Fig.~\ref{fig_qdot} the frequency is below the threshold for $\Gamma_\text{R}$ in a finite interval close to $\phi=0$, where consequently $\dot{q}=0$. 

\section{Charge imbalance voltage estimation}
We propose a measurement of the charge imbalance due to an N-S junction close to the constriction that creates a net charge $\dot{q}$, as defined in Eq.~(5) in the main text. Following the lines of Ref.~\cite{Tinkham1972b}, the normal metal and superconductor are connected via a tunnel junction with the conductance $G_T$. The current leaving the normal metal due to the finite voltage can be given as
\begin{equation}
I_V=\frac{G_T}{e}\int_\Delta^{\infty}dE\frac{E}{\sqrt{E^2-\Delta^2}}\left[f(E+eV)-f(E-eV)\right]\ ,
\end{equation}
where $V$ is the voltage across the N-S junction and $f$ is the Fermi distribution. In the low voltage limit, $eV\ll T$, one can approximate $I_V\approx-G_T \frac{V}{T}\frac{c_0}{\nu_0}$ with $c_0=\nu_0 \sqrt{2\pi \Delta T}e^{-\Delta/T}$. At higher voltages, $T\ll eV\ll\Delta$, one may approximate $I_V\approx-\frac{1}{2e}G_T e^{eV/T}\frac{c_0}{\nu_0}$. The current entering the normal metal due to the polarization of quasiparticles is given as
\begin{equation}
I_q=\frac{G_T}{e}\int_\Delta^{\infty}dE\left[f^q_\text{e}(E)-f^q_\text{h}(E)\right]\ .
\end{equation}
where $f^q_\text{e,h}$ are the nonequilibrium distributions of the electron- and hole-like quasiparticles in the lead. For the purpose of this estimate it is sufficient to express this current contribution simply as $I_q=\frac{G_T}{e}\frac{\rho}{\nu_0}$, where $\rho$ represents the non-equilibrium density of quasiparticles.

In order to estimate $\rho$ in terms of the charge transfer $\dot{q}$, we can apply a simple diffusion model. Consider the steady state diffusion equation for the spatially resolved charge density $\rho(\vec{x})$, $\vec{x}=(x,y,z)$,
\begin{equation}\label{eq_diffusion}
D\vec{\nabla}_{\vec{x}}^2\rho(\vec{x})-\tau^{-1}_q\rho(\vec{x})=-\dot{q}\delta(\vec{x})+\frac{I_\text{out}}{e}\delta(\vec{x}-\vec{x}_T)\ ,
\end{equation}
where $D$ is the quasiparticle diffusion coefficient, the charge imbalance source $\sim\dot{q}$ is placed at the axis origin, and $I_\text{out}=I_V+I_q$ is the net current leaving through the tunnel detector situated at $\vec{x}_T$. We summarize the relaxation processes for the nonequilibrium quasiparticle density (as mentioned in the main text) in a single rate $\tau_q^{-1}$. Furthermore, we impose a hard wall boundary condition at the constriction.

We consider two possibilities to probe the charge imbalance. The first consists of a voltage probe, where the voltage across the N-S junction is set to $V=V_\text{out}$ such that the net current at the junction cancels, $I_\text{out}=0$. This voltage is directly sensitive to $\rho$, i.e., $eV_\text{out}=T\rho/c_0$ ($eV\ll T$) or $eV_\text{out}\simeq T \ln(\rho/c_0)$ ($eV\gg T$). Due to the condition $I_\text{out}=0$, the drain term on the right-hand side of Eq.~\eqref{eq_diffusion} is zero, and we find that the density $\rho(x)$ simply decays as $\sim e^{-x/L_q}$ with the charge imbalance decay length $L_q=\sqrt{D\tau_q}$. In order to provide an estimate independent of the geometric details, we simply average $\rho(x)$ over this length scale (a good approximation as long as the voltage probe is situated within $L_q$) and we find $\rho\simeq\tau_q\dot{q}/\mathcal{V}$ where $\mathcal{V}$ is the volume of the lead at the length scale $L_q$. Thus one recovers the estimate of the voltage $V_\text{out}$ as in Eq.~(6) in the main text.

Alternatively, the charge imbalance may be measured by a direct current probe where the N-S junction is grounded. Hence, $I_V=0$ and $I_\text{out}=I_q=\frac{G_T}{e\nu_0}\rho(\vec{x}_T)$, which means that the probe affects the nonequilibrium density $\rho$. Solving the diffusion equation~\eqref{eq_diffusion} then readily provides $I_\text{out}$ in terms of the net charge transfer $\dot{q}$. In the limit when the distance between source and drain is $\ll L_q$ and $G_T\gg R_q^{-1}$ (where $R_q^{-1}\simeq e^2D\nu_0\mathcal{V}/L_q^2$) one obtains $I_\text{out}\approx e\dot{q}$. Therefore if the current probe is close enough, a high conductance allows ideally for a detection of the full charge transfer $\dot{q}$.

\bibliographystyle{apsrev}
\bibliography{bib_superconducting}